\documentclass{aa}  
\makeatletter
\renewcommand*\aa@pageof{, page \thepage{} of \pageref*{LastPage}}
\makeatother
\usepackage{adjustbox}
\usepackage{graphicx}
\usepackage{txfonts}
\usepackage{hyperref}
\usepackage{color}

\begin{document} 

    \titlerunning{The bright long-lived Type~II SN~2021irp powered by aspherical CSM interaction (II)}
    \authorrunning{T. M. Reynolds et al.}

   \title{The bright long-lived Type~II SN~2021irp powered by aspherical circumstellar material interaction (II): Estimating the CSM mass and geometry with polarimetry and light curve modeling}

   \author{T. M. Reynolds
          \inst{1,2,3}
          \and
          T. Nagao\inst{1,4,5}
         \and
         K. Maeda\inst{6}
         \and
         N. Elias-Rosa\inst{7,8}
        \and
         M. Fraser\inst{9}
          \and
         C. Guti\'errez\inst{10,8}
         \and
         T. Kangas\inst{11,1}
         \and
         H. Kuncarayakti\inst{1}
         \and
         S. Mattila\inst{1,12}
         \and
         P. J. Pessi\inst{13}
          }

   \institute{
            Tuorla Observatory, Department of Physics and Astronomy, University of Turku, FI-20014 Turku, Finland
            \and
            Cosmic Dawn Center (DAWN)
            \and
            Niels Bohr Institute, University of Copenhagen, Jagtvej 128, 2200 København N, Denmark
            \and
            Aalto University Mets\"ahovi Radio Observatory, Mets\"ahovintie 114, 02540 Kylm\"al\"a, Finland
            \and
            Aalto University Department of Electronics and Nanoengineering, P.O. BOX 15500, FI-00076 AALTO, Finland
            \and
            Department of Astronomy, Kyoto University, Kitashirakawa-Oiwake-cho, Sakyo-ku, Kyoto 606-8502, Japan
            \and
            INAF Osservatorio Astronomico di Padova, Vicolo dell’Osservatorio 5, 35122 Padova, Italy
            \and
            Institute of Space Sciences (ICE, CSIC), Campus UAB, Carrer de Can Magrans, s/n, E-08193 Barcelona, Spain
            \and
            UCD School of Physics, L.M.I. Main Building, Beech Hill Road, Dublin 4,D04 P7W1, Ireland
            \and
            Institut d'Estudis Espacials de Catalunya (IEEC), Edifici RDIT, Campus UPC, 08860 Castelldefels (Barcelona), Spain
            \and
            Finnish Centre for Astronomy with ESO (FINCA), FI-20014 University of Turku, Finland
             \and
             School of Sciences, European University Cyprus, Diogenes Street, Engomi, 1516, Nicosia, Cyprus
            \and 
            The Oskar Klein Centre, Department of Astronomy, Stockholm University, Albanova University Center, SE 106 91 Stockholm, Sweden
             }

   \date{...}

 
  \abstract
   {There is evidence for interaction between supernova (SN) ejecta and massive circumstellar material (CSM) in various types of SNe. The mass-ejection mechanisms that produce massive CSM are unclear, and studying interacting SNe and their CSM can shed light on these mechanisms and the final stages of stellar evolution.}
   {We aim to study the properties of the CSM in the bright, long-lived hydrogen-rich (Type~II) SN~2021irp, which is interacting with a massive aspherical CSM.}
   {We present imaging- and spectro-polarimetric observations of SN~2021irp. By modelling its polarisation and bolometric light curve, we derive the mass and distribution of the CSM.
   }
   {SN~2021irp shows a high intrinsic polarisation of $\sim$0.8\%. This high continuum polarisation suggests an aspherical photosphere created by an aspherical CSM interaction. Based on the bolometric light curve evolution and the high polarization, SN~2021irp can be explained as a typical Type~II SN interacting with a CSM disk with a corresponding mass-loss rate and half-opening angle of $\sim0.035$ - $0.1$ M$_{\odot}$ yr$^{-1}$ and $\sim30$ - $50$\textdegree, respectively. The total CSM mass derived is $\gtrsim 2$ M$_{\odot}$.
   We suggest that this CSM disk was created by some process related to binary interaction, and that SN~2021irp is the end product of a typical massive star (i.e. with Zero-Age-Main-Sequence mass of $\sim 8-18$ M$_{\odot}$) that has a separation and/or mass ratio with its companion star that led to an extreme mass ejection within decades of explosion.
   Based on the observational properties of SN~2021irp and similar SNe, we propose a general picture for the spectroscopic properties of Type II SNe interacting with a massive disk CSM. 
   }
   {}


   \maketitle
%

\section{Introduction}

Evidence for massive circumstellar material (CSM) in the vicinity of the progenitors of hydrogen-rich (Type~II) core-collapse supernovae (SNe) has been accumulating over the last decades, indicating the presence of extensive mass loss immediately before the SN explosions \citep[$\lesssim$ thousands of years before explosion; e.g.,][]{Smith2017, Nyholm2020, Fraser2020}. The most common subclass of Type II SNe is known as Type~IIP, for the plateau phase in their light curves \citep{Barbon1979}, and are associated with explosions of red supergiant stars with Zero-Age-Main-Sequence masses between $\sim8$ and $\sim$18 $M_{\odot}$ \citep[see e.g.][]{VanDyk2003,Smartt2004,Smartt2015a}; although see e.g. \citet{Beasor2024} who argue for somewhat higher masses. These SNe are now believed to generally have nearby CSM, suggested by short-lived narrow emission lines in their early-phase spectra \citep[e.g.,][]{Khazov2016,Yaron2017,Boian2020,Bruch2021} and also by rapid rises in their early light curves \citep[e.g.,][]{GonzalezGaitan2015,Forster2018}. The Type IIL SN subclass, which exhibits a linear decline in magnitude rather than a plateau, also shows signs of significant CSM interaction \citep[e.g.,][]{Morozova2017, Morozova2018, Maeda2023}. The mass-loss mechanism/s responsible for the CSM have not yet been established, although some mechanisms have been proposed \citep[e.g.,][]{Humphreys1994,Langer1999,Yoon2010,Arnett2011,Chevalier2012,Quataert2012,Soker2013,Shiode2014,Smith2014a,Woosley2015,Quataert2016,Fuller2017}.

Type~IIn SNe, which are characterized by a blue continuum and narrow Balmer lines in their spectra \citep{Schlegel1990}, are some of the most extreme cases of interacting SNe. They are mainly powered by an interaction between the SN ejecta and the CSM \citep[e.g.,][]{Smith2017, Nyholm2020, Fraser2020}. The characteristic Balmer emission lines have narrow cores ($\sim$ ten to hundreds of km s$^{-1}$), indicating that they arise from the slow-moving material around the SN ejecta and sometimes broad Lorentzian wings produced by electron scattering occurring in the dense CSM  \citep[e.g.,][]{Schlegel1990,Chugai2001,Smith2017}. Since these narrow lines are believed to originate from gas in the unshocked CSM, which is ionised/excited by high-energy photons from the interaction shocks, they are regarded as an indicator of a major CSM interaction. The estimated total mass of the CSM required to explain the observed radiated energy can be as large as a few tens of solar masses in extreme cases \citep[e.g. SN~2006gy;][]{Smith2010}. 

Type IIn SNe show diverse properties in their light curves and spectra and are believed to originate from diverse progenitor systems \citep[e.g.,][]{Smith2017, Nyholm2020, Fraser2020}. Some SNe are luminous for only a short period ($\sim$several tens of days); while others last longer ($\sim$several years) and there is a diversity in their absolute peak magnitudes \citep[$\sim$~-17 to $\sim$~-22 mag;][]{Nyholm2020}. In addition, the time evolution of the shapes of the Balmer lines varies, showing intermediate-width/broad components in some cases \citep[e.g.,][]{Smith2017}. These diverse observational properties should reflect different properties of the SN ejecta and the CSM.

It is important to investigate the CSM geometries in interacting SNe, which are directly related to the mass-loss mechanisms and influence the observational properties. Polarimetry is a powerful tool for investigating the shape of spatially unresolved sources, such as SNe. In fact, several Type~IIn show high polarisation, implying non-spherical CSM interactions \citep[e.g.,][]{Leonard2000,Patat2011,Reilly2017,Bilinski2018,Kumar2019,Mauerhan2024,Bilinski2024}, although the observational sample is limited.

In general, it is difficult to study the properties of the embedded SNe in interaction-dominated events because the interaction features dominate over the ones from the SN ejecta. This makes it challenging to identify the progenitor systems of Type~IIn SNe. However, in the case of aspherical CSM interaction, we can sometimes extract information about the embedded SN. For example, we clearly see a Type Ia~SN interacting with hydrogen-rich CSM in the cases of Type Ia-CSM SNe \citep[e.g.,][]{Hamuy2003,Dilday2012}. The spectra of this class of SNe show narrow Balmer lines similar to those seen in other types of interacting SNe as well as the typical broad lines seen in thermonuclear explosions, such as silicon, sulphur, and iron. \citet[][]{Uno2023a,Uno2023b} posit that the ejecta of the Type~Ia-CSM SN 2020uem are visible because the CSM interaction happens in a disk-shape configuration, allowing direct observations of the SN ejecta even though it has an extensive ongoing CSM interaction similar to Type~IIn SNe, and thus the interaction shock is optically thick. In fact, they pointed out the similarity of the spectral lines in SN~2020uem to those in 91T-like Type~Ia SNe.

Photometric and spectroscopic observations of the interacting SN~2021irp are presented and analysed in a companion paper to this work, Reynolds et al. (2025), hereafter Paper I. The SN is luminous, with a peak absolute mag of $M_{o}$\footnote{The magnitude here is in the orange ($o$) band of the Asteroid Terrestrial-impact Last Alert System  \citep[ATLAS;][]{Tonry2018,Smith2020}, corresponding approximately to $r + i$.}$ < -19.4$ mag (the actual peak was missed and was likely brighter); and long-lived, remaining brighter than $M_{o} = -18$ mag for $\sim$~250~d. The total radiated energy measured was >~$2.6\times~10^{50}$ erg, significantly larger than typical for a Type II SN, so an additional power source is required. Spectra of the SN are available after $\sim$~200~d and show strong multi-component Balmer emission lines, with a broad component with Full-Width-Half-Maximum (FWHM) $~\sim~$8000 km~s$^{-1}$ and an intermediate width component of $\sim$~2000 km~s$^{-1}$, along with emission lines of He, Fe, Ca and Na. The spectra evolve slowly, but there is a dramatic erosion in the red wing of many emission lines. This is consistent with dust formation in the SN, which is supported by decline rate and color changes observed in the photometry. We conclude in Paper I that the SN is mainly powered by CSM interaction and that the CSM must exhibit significant asymmetry. 

In this work, we further investigate the CSM geometry of SN~2021irp through polarimetric observations and modelling of the bolometric light curve. In Sect. \ref{sec:obs_data} we describe the observation and reduction of our spectro-polarimetric and imaging-polarimetric data. In Sect. \ref{sec:pol_properties} we estimate the interstellar polarisation towards the SN and then measure and discuss the intrinsic SN polarisation. In Sect. \ref{sec:LC_modeling} we model the bolometric light curve of a Type II SN interacting with a disk-shaped CSM and compare the model results with observations of SN~2021irp. In Sect. \ref{sec:CSM_properties}, we discuss the CSM properties derived from our model. In Sect. \ref{sec:discussion} we discuss the implications of our modelling.
We adopt the parameters for the SN given in Paper I: the explosion epoch of 59310.3 $\pm 3$ (MJD), the total extinction of $A_V = 1.31 \pm 0.04$ mag and the redshift of z=$0.0195\pm0.0001$. All phases are given in rest-frame days with respect to the explosion epoch.


\section{Observations and data reduction}
\label{sec:obs_data}

\begin{table*}
\caption{Log of polarimetric observations of SN~2021irp}
\centering
\begin{tabular}{c c c c c c} \hline\hline 
Phase (days) & Date (UT) & MJD & Telescope & Instrument & observing mode \\
\hline
190.1 & 2021-10-17.08 & 59504.08 & NOT & ALFOSC & imaging polarimetry (R)  \\
203.0 & 2021-10-30.26 & 59517.26 & VLT & FORS2 &  spectro-polarimetry \\
261.8 & 2021-12-29.19 & 59577.19 & VLT & FORS2 &  imaging polarimetry (FILT\_815\_13)\\
\hline\hline 
\end{tabular}
\label{tab:pol_log}
\end{table*}

   \begin{figure*}
   \centering
   \includegraphics[width=\columnwidth]{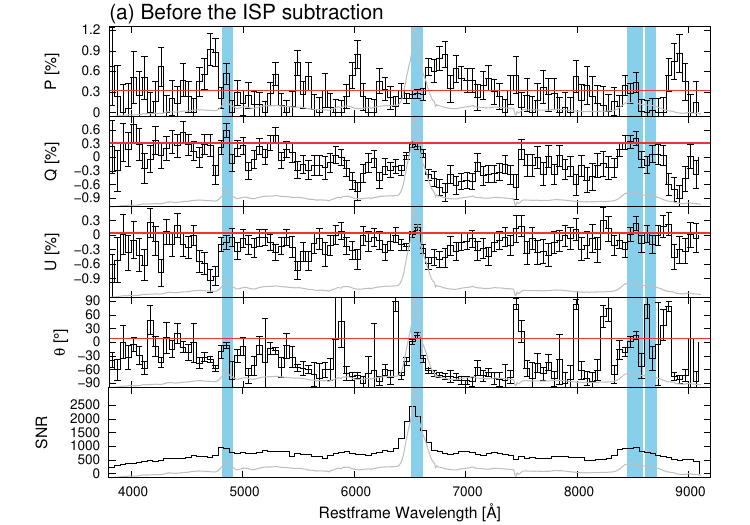}
   \includegraphics[width=\columnwidth]{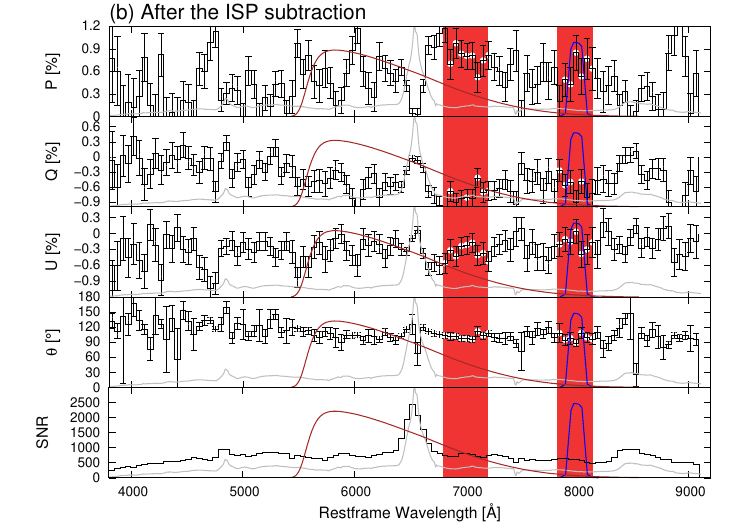}
      \caption{
      Polarisation spectra of SN 2021irp before and after the ISP subtraction. (a) Total polarisation $P$, Stokes parameters $Q$ and $U$, polarisation angle $\theta$, and signal-to-noise ratio (SNR) before the ISP subtraction at a phase of 203.0 d (black lines). The data are binned to 50 {\AA} per point. The grey lines in the background of each plot are the unbinned total-flux spectra at the same epoch. 
      The ISP is described by $P_{\rm{ISP}}=0.32$ \%, $\theta_{\rm{ISP}} = 8.8$\textdegree~ (red lines). The blue hatching shows the adopted wavelength range for the ISP-dominated components. (b) Same as (a), but after the ISP subtraction. The brown and blue lines show the transmission curves of the R band and FILT\_815\_13. The red hatching shows the adopted wavelength range for the estimate of the continuum polarisation.
              }
         \label{fig:specpol}
   \end{figure*}

We have conducted spectro-polarimetric and imaging-polarimetric observations of SN~2021irp using the FOcal Reducer/low-dispersion Spectrograph 2 (FORS2; Appenzeller et al. 1998) mounted on the Very Large Telescope (VLT) and the Alhambra Faint Object Spectrograph and Camera (ALFOSC) on the Nordic Optical Telescope (NOT). The log of the observations is shown in Table~\ref{tab:pol_log}.

The set-ups of the spectro-polarimetric and imaging-polarimetric observations and the analysis are similar to those reported by \citet[][]{Nagao2024a,Nagao2024b}. For the spectro-polarimetric observation by FORS2/VLT, the spectrum produced by a grism is split by a Wollaston prism into two beams with orthogonal direction of polarisation (ordinary (o) and extraordinary (e) beams) after passing through a half-wave retarder plate (HWP). We adopted 4 HWP angles ($0^{\circ}$, $22.5^{\circ}$, $45^{\circ}$ and $67.5^{\circ}$). We used the low-resolution G300V grism and a $1.0$ arcsec slit, giving a spectral coverage of $3800-9200$ {\AA}, a dispersion of $\sim 3.2$ {\AA}~pixel$^{-1}$ and a resolution of $\sim 11.5$ {\AA} (FWHM) at $5580$ {\AA}. For the imaging-polarimetric observations with FORS2/VLT, the same instrumental set-up was adopted, with a narrow-band filter (FILT\_815\_13) instead of the grism in the optical path. For the imaging polarimetry with ALFOSC/NOT, the same instrument set-up was adopted but with the R-band filter.

The data were reduced with IRAF\citep[][]{Tody1986,Tody1993} using standard methods as described, e.g., in \citet[][]{Patat2006}. The ordinary and extraordinary beams were extracted by the PyRAF \texttt{apextract.apall} task with a fixed aperture size of $10$ pixels and then separately binned in 50 {\AA} bins in order to improve the signal-to-noise ratio. The HWP zeropoint angle chromatism was corrected based on the data in the FORS2 user manual \footnote{\url{http://www.eso.org/sci/facilities/paranal/instruments/fors/doc/VLT-MAN-ESO-13100-1543_P07.pdf}}. The wavelength scale for the Stokes parameters was corrected to the rest-frame using the redshift of the host galaxy.
As for the imaging-polarimetric data, the counts for the ordinary and extraordinary beams were measured using aperture photometry after the bias and flat corrections. For the aperture photometry, we adopted an aperture whose size is twice as large as the FWHM of the ordinary beam’s point-spread function and a ring background area whose inner and outer radii are three and four times as large as the FWHM, respectively. From the measurements, we calculated the Stokes parameters, the linear polarisation degree and the polarisation angle.
The polarisation bias in the spectro- and imaging-polarimetric data was subtracted using the standard method described in \citet{Wang1997}.

\section{Polarimetric properties} \label{sec:pol_properties}

\subsection{Interstellar polarisation estimation}

To estimate the interstellar polarisation (ISP), we use a frequently adopted method based on the assumption that the emission peaks of the strong spectral lines purely reflect the ISP due to the depolarisation of the intrinsic SN polarisation by the multiple scattering and/or absorption 
\citep[see][and references therein]{Nagao2023}.
We extract the signals of the H$\beta$ $\lambda 4861$, H$\alpha$ $\lambda 6563$ and Ca II triplet $\lambda 8498/8542/8662$ lines with a 100 {\AA} window around their peaks using the spectro-polarimetric data (see the blue hatching in Figure~\ref{fig:specpol}a). By taking average values of these signals, we estimate the polarisation degree and angle of the ISP as $P_{\rm{ISP}}=0.32$ \% and $\theta_{\rm{ISP}} = 8.8$\textdegree.
This polarisation degree is consistent with the empirical relation between the extinction and the consequent polarisation ($P \lesssim 9 E(B-V)$; Serkowski et al. 1975), which produces $P_{\rm{ISP}} \lesssim 3.8$ \% towards the assumed extinction ($E(B-V)=0.42$ mag; see Paper~I).

Based on the polarisation degree, the ISP can, in principle, originate from dust either in the Milky Way (MW) or in the host, according to the above empirical relation. However, it might be reasonable to consider the origin as MW dust because the extinction in the MW is higher than that in the host by an order of magnitude. The polarisation angle may support this interpretation, although it is not definitive. Based on the 3D dust map \citep[][]{Green2019}, the majority of the dust responsible for the MW extinction is located around 380-400 pc away from us. The polarisation angles of field stars at similar locations are relatively similar to the observed angle for the ISP \citep[$\sim 0$\textdegree;][]{Mathewson1970}. In any case, this speculation on the origin of the ISP does not affect our results below.

\subsection{Intrinsic supernova polarisation}

Figure~\ref{fig:specpol}b shows the polarisation spectrum after the ISP subtraction. It shows relatively constant polarisation angles over the observed wavelength region with high polarisation degrees ($\sim 0.8$ \%) at wavelength regions without strong lines and low degrees at strong emission lines and at bluer wavelengths where there are many weak blended lines. These polarisation properties can be interpreted as a high continuum polarisation whose degrees and angles are constant through wavelengths and depolarized at wavelengths of emission lines by the multiple scattering and/or absorption.

For the estimation of the continuum polarisation, we averaged the signals of the polarisation spectrum in the wavelength regions from 6800 {\AA} to 7200 {\AA} and from 7820 {\AA} to 8140 {\AA} \citep[see the red hatching in Figure~\ref{fig:specpol}b; e.g.,][]{Chornock2010, Nagao2019, Nagao2021, Nagao2024a}, after the ISP subtraction. The derived continuum polarisation at 203.0~d is as follows: $P=0.77 \pm 0.06$, $\theta=100.5$\textdegree$~\pm~ 2.5$\textdegree. 
The values from imaging polarimetry at different epochs are consistent with those from spectro-polarimetry, all showing a $\sim 1$ \% level of polarisation with an angle of $\sim 100$\textdegree. The $R$-band polarisation at 190.1~d is $P=0.88 \pm 0.58,~\theta=110.2$\textdegree~$\pm$ 18.8\textdegree, while the polarisation through the FILT\_815\_13 filter at 261.8~d is $P=1.47 \pm 0.40$, $\theta=116.8$\textdegree ~$\pm$ 8.7\textdegree.

The high continuum polarisation in SN~2021irp, which is mainly powered by CSM interaction (see Paper~I), implies an aspherical electron-scattering-dominated photosphere and thus an aspherical CSM interaction.
This high polarisation degree ($\sim 0.8$ \%) corresponds to a photosphere of an oblate ellipsoid with axis ratios of 0.8, 0.7, 0.5 and 0.2 towards viewing angles $\theta_{\rm{obs}}$ of $\sim 90$\textdegree, $\sim 60$\textdegree~, $\sim 40$\textdegree~ and $\sim 30$\textdegree~from the polar direction, respectively, based on the electron-scattering atmosphere model by \citet[][see also \citet{Uno2023b} for more details]{Hoflich1991}. This suggests that the CSM structure should be aspherical, e.g., with a disk, jet or clumpy geometry. The constant polarisation angles from $\sim 190$ to $\sim 260$~d imply that the aspherical structure of the scattering regions remains constant at least during this period.
As a approximation, we assume that the asphericity of the photosphere in a SN interacting with disk-like CSM with a half-opening angle of $\theta_{0}$ is similar to that of an oblate ellipsoid with axis ratio of $\tan(\theta_{0})/2$, although the actual shape of the photosphere is more complicated (see Section~\ref{sec:CSM_properties}). In order to get a high polarization degree of $\gtrsim 0.8$ \%, the values of  $\theta_{0}$ should be $\lesssim 58$\textdegree~ , $\lesssim 55$\textdegree~ , $\lesssim 45$\textdegree~  and $\lesssim 22$\textdegree~  for $\theta_{\rm{obs}}$ of $\sim 90$\textdegree, $\sim 60$\textdegree~, $\sim 40$\textdegree~ and $\sim 30$\textdegree~, respectively \citep[][see also \citet{Uno2023b} for more details]{Hoflich1991}. Additionally, since the viewing angle of SN~2021irp should be a CSM-free direction (see Paper~I), the viewing angle should satisfy the following condition: $\theta_{\rm{obs}} \lesssim 90$\textdegree$-~ \theta_{0}$. Taking these conditions into account, we can rule out very large opening angles for the disk, and roughly conclude that $\theta_{0} \lesssim 50$\textdegree~. Further discussion on the CSM geometry can be found in Section~\ref{sec:CSM_properties}.

\begin{figure}
   \centering
   \includegraphics[width=\columnwidth]{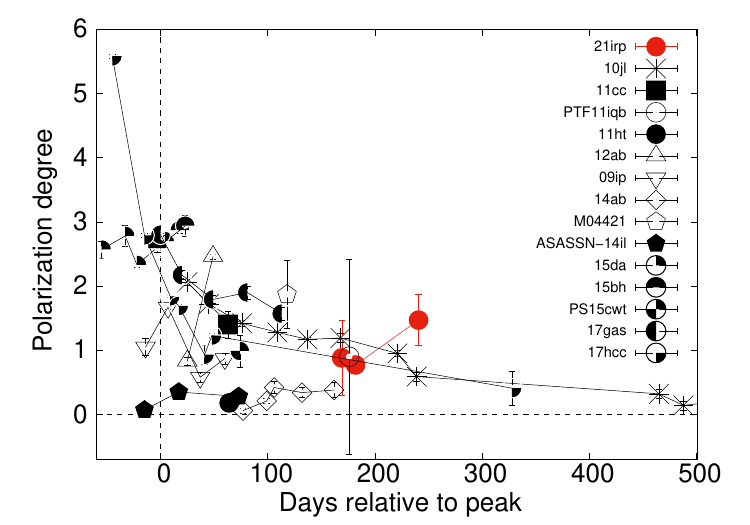}
      \caption{Time evolution of the continuum polarisation in SN~2021irp, compared to those of Type IIn SNe in \citet{Bilinski2024}. Here, the peak date for SN~2021irp is set to the date of the maximum magnitude in the ATLAS-o-band light curve as 59331.3 (MJD), although the actual peak was not caught by the observations.
              }
         \label{fig:comp_pol}
   \end{figure}

In Fig. \ref{fig:comp_pol}, we show the evolution of the continuum polarisation in SN~2021irp compared to other Type IIn SNe, as collated in \citet{Bilinski2024}. The sample of SNe with detected polarisation at such late phases is small. The polarisation observed in SN~2021irp is similar to that observed in SN~2010jl and SN~2017hcc, assuming linear evolution of the polarisation between the observations at $\sim$ 50~d and $\sim$ 350~d for SN~2017hcc. We do not have observations of SN~2021irp at early times, but SN~2017hcc was observed to have an extremely large $\sim$ 6\% polarisation at pre-peak \citep{Mauerhan2024}.

\section{Light curve modelling} \label{sec:LC_modeling}

We calculate the bolometric luminosity from the interaction shock between the SN ejecta and a disk-shaped CSM ($L_{\rm{disk}}$) using the light curve model presented in \citet[][]{Uno2023b}, which is a modified version of the models by \citet[][]{Moriya2013,Nagao2020}.
We note that the explosion date for SN~2021irp is well constrained by observations, with an uncertainty of $\sim 3$~d, despite the fact that the majority of the first $\sim 100$ days of the SN were not observed (see Paper I for details).

\subsection{Supernova ejecta properties}
For simplicity, we assume typical ejecta properties for a Type II SN as follows:
the ejecta mass: $M_{\rm{ej}}=10$ M$_{\odot}$ and the kinetic energy of the ejecta: $E_{\rm{ej}}=10^{51}$ erg.

For the density profile of the ejecta, we adopt the double power-law distribution \citep[e.g.,][]{Moriya2013}, which is derived from hydrodynamics simulations \citep[e.g.,][]{Matzner1999}:
\begin{eqnarray}
\label{eq:rho_ej}
\widetilde{\rho}_{\rm{ej}} (v_{\rm{ej}},t) = 
\left\{
    \begin{array}{l}
      \frac{1}{4 \pi (n-\delta)} \frac{[2(5-\delta)(n-5) E_{\rm{ej}}]^{\frac{n-3}{2}}}{[(3-\delta)(n-3) M_{\rm{ej}}]^{\frac{n-5}{2}}} t^{-3} v_{\rm{ej}}^{-n} \;\;\;\; (v_{\rm{ej}} > v_{t})\\
      \frac{1}{4 \pi (n-\delta)} \frac{[2(5-\delta)(n-5) E_{\rm{ej}}]^{\frac{\delta-3}{2}}}{[(3-\delta)(n-3) M_{\rm{ej}}]^{\frac{\delta-5}{2}}} t^{-3} v_{\rm{ej}}^{-\delta} \;\;\;\; (v_{\rm{ej}} < v_{t})
    \end{array}
  ,\right.
\end{eqnarray}
where $v_{\rm{ej}} (r,t)$ ($= r/t$) is the SN ejecta velocity at radius, $r$, and time, $t$, and 
\begin{equation}
    v_{t} = \left[ \frac{2(5-\delta)(n-5)E_{\rm{ej}}}{(3-\delta)(n-3)M_{\rm{ej}}} \right]^{\frac{1}{2}}.
\end{equation}
Hereafter, we use $\rho_{\rm{ej}} (r,t)$, which has the independent variable $r$ instead of $v_{\rm{ej}}$ and is given by $\widetilde{\rho}_{\rm{ej}} (v_{\rm{ej}},t)$.
Here, we assume that $n=12$ and $\delta = 1$, which are typical values for Type II SNe \citep[e.g.,][]{Matzner1999}.\\

\subsection{Circumstellar material properties}

As an initial distribution of the CSM, we adopt a disk structure with a half-opening angle, $\theta_{0}$. 
We use the radial distribution of the CSM as 
\begin{equation}
\rho_{\rm{csm}}(r) = \frac{\widetilde{\dot{M}}_{\rm{csm}}}{4\pi v_{\rm{csm}}} r^{-2} = \frac{\dot{M}_{\rm{disk}}}{4\pi v_{\rm{csm}} \omega_{\rm{disk}}} r^{-2} = D r^{-2}.
\end{equation}
The isotropic mass-loss rate ($\widetilde{\dot{M}}_{\rm{csm}}$) is linked with the actual mass-loss rate ($\dot{M}_{\rm{disk}}$) for the disk-shaped CSM as follows: $\widetilde{\dot{M}}_{\rm{csm}}= (4 \pi \dot{M}_{\rm{disk}})/\Omega_{\rm{disk}} = \dot{M}_{\rm{disk}}/\omega_{\rm{disk}}$. Here, $\Omega_{\rm{disk}}$ is the solid angle of the CSM disk, and $\omega_{\rm{disk}} = \Omega_{\rm{disk}}/4 \pi = \sin{\theta_{0}}$.
The actual mass-loss rate ($\dot{M}_{\rm{disk}}$) expresses the total mass that is ejected per unit time. Thus, for the same value of $\dot{M}_{\rm{disk}}$, the density of the CSM, which controls the evolution of the CSM interaction, depends on the half-opening angle of the disk. On the contrary, the isotropic mass-loss rate ($\widetilde{\dot{M}}_{\rm{csm}}$) is a good indicator to understand the density of the CSM and thus the evolution of the interaction shock.
This distribution with $r^{-2}$ is expected for a steady mass loss from a progenitor system with a constant mass-loss rate of $\dot{M}_{\rm{disk}}$ and wind velocity of $v_{\rm{csm}}$.
The CSM velocity is set to $v_{\rm{csm}} = 50$ km s$^{-1}$, which is consistent with the value estimated for SN~2021irp from the high-dispersion spectroscopic observation ($\lesssim 85$ km s$^{-1}$; see Paper~I). The location of the inner edge of the disk is assumed to correspond to the progenitor radius before the explosion: $R_{\rm{csm}} = R_{\rm{p}}$. In the outer region, the CSM is assumed to extend to infinity. 

\subsection{Circumstellar material interaction}
We calculate the evolution of the shocked shell from the equation of motion of the shock, assuming the shell is physically-thin compared to its radius \citep[see][for details]{Uno2023b}.
\begin{eqnarray}
\label{eq:eom}
    M_{\rm{sh}} (t) \frac{dv_{\rm{sh}}(t)}{dt} &=& 4 \pi r_{\rm{sh}}^{2}(t) \Bigl[ \rho_{\rm{ej}} (r_{\rm{sh}}(t),t) \left( v_{\rm{ej}}(r_{\rm{sh}}(t),t)-v_{\rm{sh}}(t) \right)^{2} \nonumber\\
    && - \rho_{\rm{csm}} (r_{\rm{sh}}(t)) \left( v_{\rm{sh}}(t)-v_{\rm{csm}} \right)^{2} \Bigr],
\end{eqnarray}
where $M_{\rm{sh}}(t)$ is the total mass of the shocked SN ejecta and CSM and $v_{\rm{sh}}(t)$ is the velocity of the shocked shell at a given time, $t$. Here $M_{\rm{sh}}(t)$ is expressed as follows:
\begin{equation}
    M_{\rm{sh}} (t) = \int_{R_{\rm{csm}}}^{r_{\rm{sh}}(t)} 4 \pi r^{2} \rho_{\rm{csm}} (r) dr + \int_{r_{\rm{sh}}(t)}^{r_{\rm{ej,max}}(t)} 4 \pi r^{2} \rho_{\rm{ej}} (r,t) dr
\end{equation}
where $r_{\rm{ej,max}}(t) = v_{\rm{ej,max}} t$, and $v_{\rm{ej,max}}$ is the original velocity of the outermost layer of the SN ejecta before the interaction. Here, we assume $r_{\rm{ej,max}}(t) >> r_{\rm{sh}}(t)$ at all times. We obtain the values of $r_{\rm{sh}}(t)$ and $v_{\rm{sh}}(t)$ by numerically solving Eq.~\eqref{eq:eom}.

\subsection{Bolometric luminosity}

We calculate the bolometric luminosity from the interaction between the assumed SN ejecta and CSM, following the analytical model by \citet[][]{Uno2023b}.

\begin{equation}
\label{eq:Ldisk}
    L_{\rm{disk}} (t,\dot{M}_{\rm{disk}}) = \omega_{\rm{disk}} L_{\rm{sphere}} \left(t,\frac{\dot{M}_{\rm{disk}}}{\omega_{\rm{disk}}} \right).
\end{equation}

\subsection{Comparison of the light curve models with observations}

\begin{figure}
   \centering
   \includegraphics[width=0.5\textwidth]{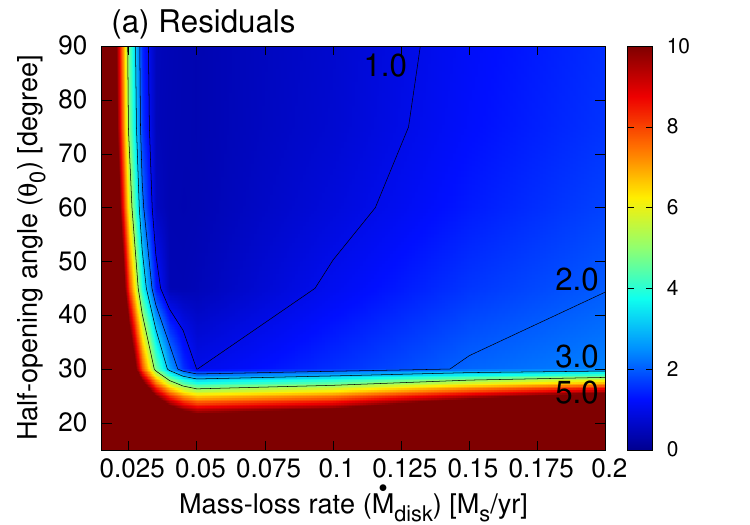}
   \includegraphics[width=0.5\textwidth]{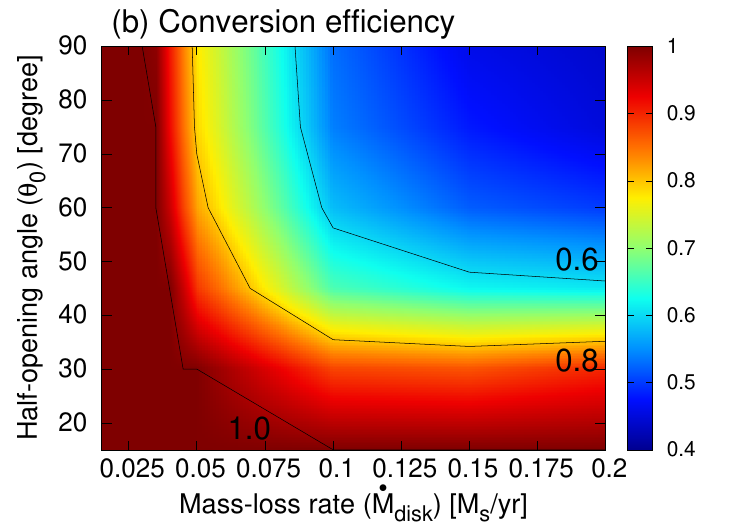}
      \caption{Results for the light curve fitting. (a): Residual between the observed pseudo-bolometric light curve and the best-fit model for each combination of the mass-loss rate ($\dot{M}_{\rm{disk}}$) and the half-opening angle ($\theta_{0}$). (b): Energy conversion efficiency ($\epsilon$) for each best-fit model.
              }
         \label{fig:best-fit}
   \end{figure}

We derived best-fit light curve (LC) models for the pseudo-bolometric light curve of SN~2021irp, which is derived from blackbody fitting using the optical photometry (see Paper~I), with the mass-loss rate ($\dot{M}_{\rm{disk}}$), the opening angle ($\theta$), and the conversion efficiency ($\epsilon$) as free parameters. We investigated the ranges of these parameters as follows: $0.015 \leqq \dot{M}_{\rm{disk}} \leqq 0.2$ M$_{\odot}$ yr$^{-1}$, $15 \leqq \theta \leqq 90$ degrees and $0.1 \leqq \epsilon \leqq 1.0$.

Figure~\ref{fig:best-fit} shows the residuals between the observations and the best-fit models for all combinations of $\dot{M}_{\rm{disk}}$ and $\theta$ and the energy conversion efficiency for each best-fit model. 
Here, the residuals are defined as in \citet[][]{Uno2023b}: $\Sigma[(\rm{model} - \rm{data})/\rm{data}]^{2}$. It is noted that we did not use the late-phase data ($>200$~d after the explosion) in the light curve fitting, in order to avoid the part with the accelerated decay (see Paper~I for the discussions on possible origins of this decay).
The total number of the data points of the pseudo-bolometric light curve that we used for the fitting ($< 200$~d after the explosion) is 94. Thus, the residual per data point is derived by dividing the above residual by 94.
According to the obtained residuals, the realistic values for the mass-loss rate and the half-opening angle (from the models with residuals $\lesssim 1.0$) are in the ranges of $0.035 \lesssim \dot{M}_{\rm{disk}} \lesssim0.1$ M$_{\odot}$ yr$^{-1}$ and $30 \lesssim \theta_{0} \lesssim90$ degrees, respectively. In a dense CSM interaction, the conversion from the kinetic to thermal energy becomes effective, where the conversion efficiency is close to unity \citep[e.g.,][]{Maeda2022}. The best-fit models show high values of the conversion efficiency from $\sim 0.5$ to 1.0 (Figure~\ref{fig:best-fit}).

\section{Derived circumstellar material properties}
\label{sec:CSM_properties}

\begin{figure}
   \centering

    \includegraphics[width=0.5\textwidth]{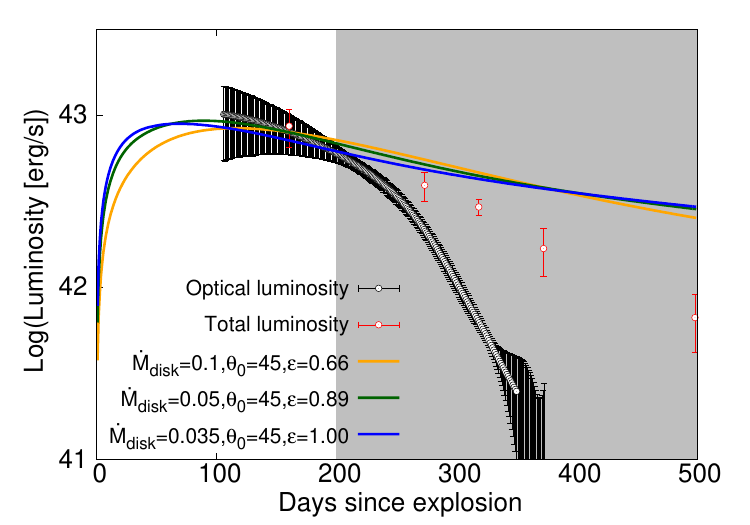}
    \caption{Time evolution of the luminosity for several well-fitting models.
    The pseudo-bolometric luminosity estimated with the optical photometry is shown with black open circles, while the total luminosity from the optical and infrared photometry is with red open circles.   
    Model parameters are listed in the figure. We excluded the late-phase data ($t\gtrsim 200$~d; gray hatching) from the light curve fitting.
              }
    \label{fig:LC_fitting_bol}
\end{figure}

\begin{figure*}
   \includegraphics[width=1\columnwidth]{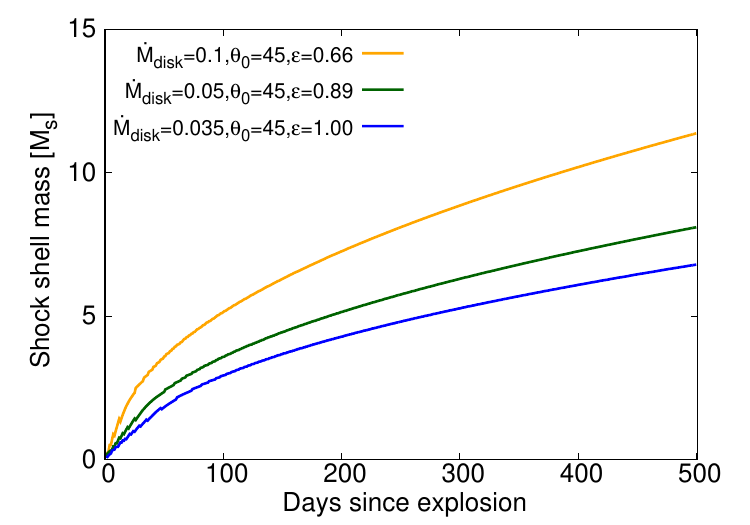}
   \includegraphics[width=1\columnwidth]{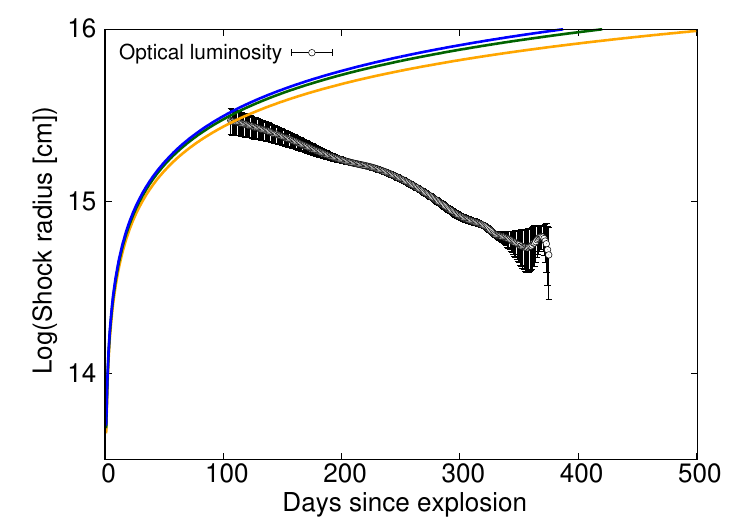}
   \includegraphics[width=1\columnwidth]{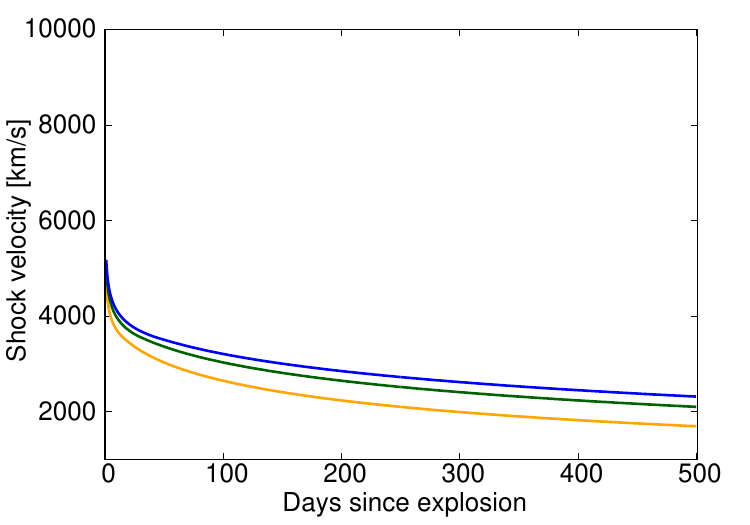}
   \includegraphics[width=1\columnwidth]{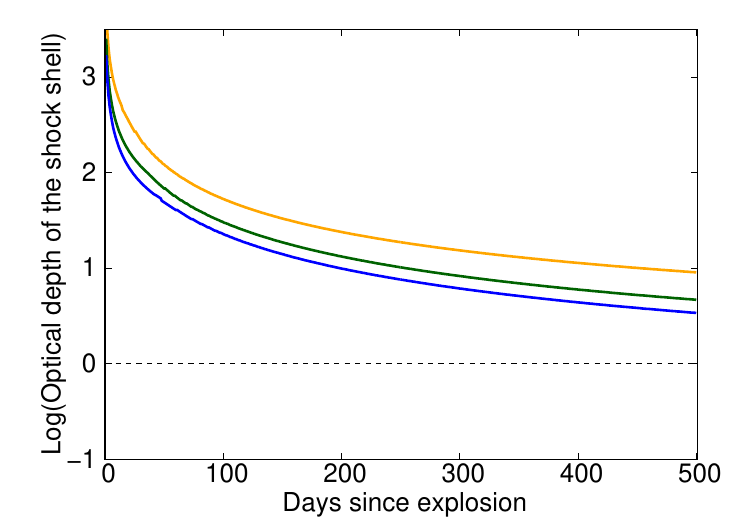}
      \caption{Time evolution of the shock shell mass (top left), shock radius (top right), shock velocity (bottom left) and optical depth of the shock shell (bottom right) for several well-fitting models. The adopted model parameters are the same in Fig. \ref{fig:LC_fitting_bol}. We excluded the late-phase data ($t\gtrsim 200$~d; grey hatching) from the light curve fitting. The black open circles in the top right panel are the blackbody radii of the optical component from the two blackbody fitting in Paper~I.
              }
         \label{fig:LC_fitting}
   \end{figure*}

Fig. ~\ref{fig:LC_fitting_bol} shows the light curves for several sets of best-fit parameters with relatively small residuals, and Fig. ~\ref{fig:LC_fitting} shows the time evolution of the mass and location of the shock shell for each parameter set.
The assumed density distribution of the CSM ($\rho \propto r^{-2}$) can roughly explain the decline rate at early phases ($< 200$~d). The steep decline of the optical luminosity after approximately 250~d is most likely caused by dust formation (see Paper~I), but even after including the IR luminosity arising from the dust, the light curve deviates from the best-fit models at late phases. This might imply that the actual CSM density distributions are slightly steeper than the assumed one ($s=2$).
The total CSM mass swept up by the interaction shock is at least several solar masses already at $\sim 200$~d after the explosion. If the assumed SN parameters are correct, the minimum value for the necessary CSM mass would be $\sim 2~M{_{\odot}}$ for the best-fit case with $\dot{M}_{\rm{disk}}=0.035$ M$_{\odot}$ yr$^{-1}$. This necessary CSM mass is within the range of the diversity in Type IIn SNe \citep[e.g.,][]{Smith2017}. 

The expected shock radii in the best-fit cases are always larger than the observed black-body radius (Figure~\ref{fig:LC_fitting}). This suggests that the CSM interaction is aspherical and the photosphere should be localized around the interaction regions. The expected shock radius is several $\times$ $10^{15}$ cm at $\sim200$~d after the explosion. Assuming a wind velocity of $\sim 50$ km s$^{-1}$, this distance corresponds to an expansion time of $\sim 10$ years, implying that the progenitor ejected $\gtrsim 2~M{_{\odot}}$ within the last $\sim 10$ years before the explosion.
In addition, the shock velocities in the best-fit cases ($\sim 3000$ km s$^{-1}$ at a few hundred days after the explosion) are much lower than the observed velocities in the Balmer lines ($\sim 8000$ km s$^{-1}$; see Paper~I). This suggests that the emitting regions for the broadest part of the Balmer lines and possibly the photospheric radiation should be in the H-rich SN ejecta rather than in the shock shell. Namely, we are looking at parts of the H-rich SN ejecta heated by the aspherical CSM interaction. 
In fact, the optical depth of the shock shell from the best-fit models is much higher than unity at the phase of $\sim 200$~d, when we already observe broad Balmer lines in the spectra of SN~2021irp (Figure~\ref{fig:LC_fitting}). This is also true for the spherical models with $\theta_0=90$\textdegree. Therefore, the CSM interaction is required to be aspherical so that we can see the SN ejecta where the broad Balmer lines arise at this phase of the SN when the shocked shell is optically thick.

Our light curve models demonstrate that the bolometric light curve can be explained by a normal Type~II SN (an explosion of an red supergiant with ejecta mass $\sim10$ M$_{\odot}$ and explosion energy of $\sim 10^{51}$ erg) interacting with several solar masses of CSM in a disk-shape.
We can also consider other aspherical morphologies for the CSM: a jet-like morphology where the CSM has a continuous radial distribution extending from a spherical cap with a small opening angle; or a blob-like morphology, which is similar but is not continuously distributed. The light curves in the cases of jet-like or blob-like CSM, would be equivalent to the disk models with a small half-opening angle of the disk, which are not preferred by the light curve fitting (see Figure~\ref{fig:best-fit}). In order to obtain similar luminosities with the jet/blob CSM distribution, we need a denser CSM than in the disk CSM case. For example, the density scale for the interaction with jet-like CSM with a half-opening angle of $\sim 30$\textdegree~ corresponds to the disk case with $\sim 8$\textdegree.
In addition, the shapes of the broad Balmer lines are rather symmetric towards the red and blue sides before the acceleration of the light curve decline ($\lesssim 250$~d after the explosion; see Paper~I), which corresponds to the likely onset of dust formation. This disfavours the scenarios with jet- or blob-like CSM, where clear multiple peaks (for bipolar jet-like CSM) or a shifted peak to the red or blue side (for a blob-like CSM) would be expected in the observed Balmer lines.

\begin{figure*}
   \centering
   \includegraphics[width=2\columnwidth]{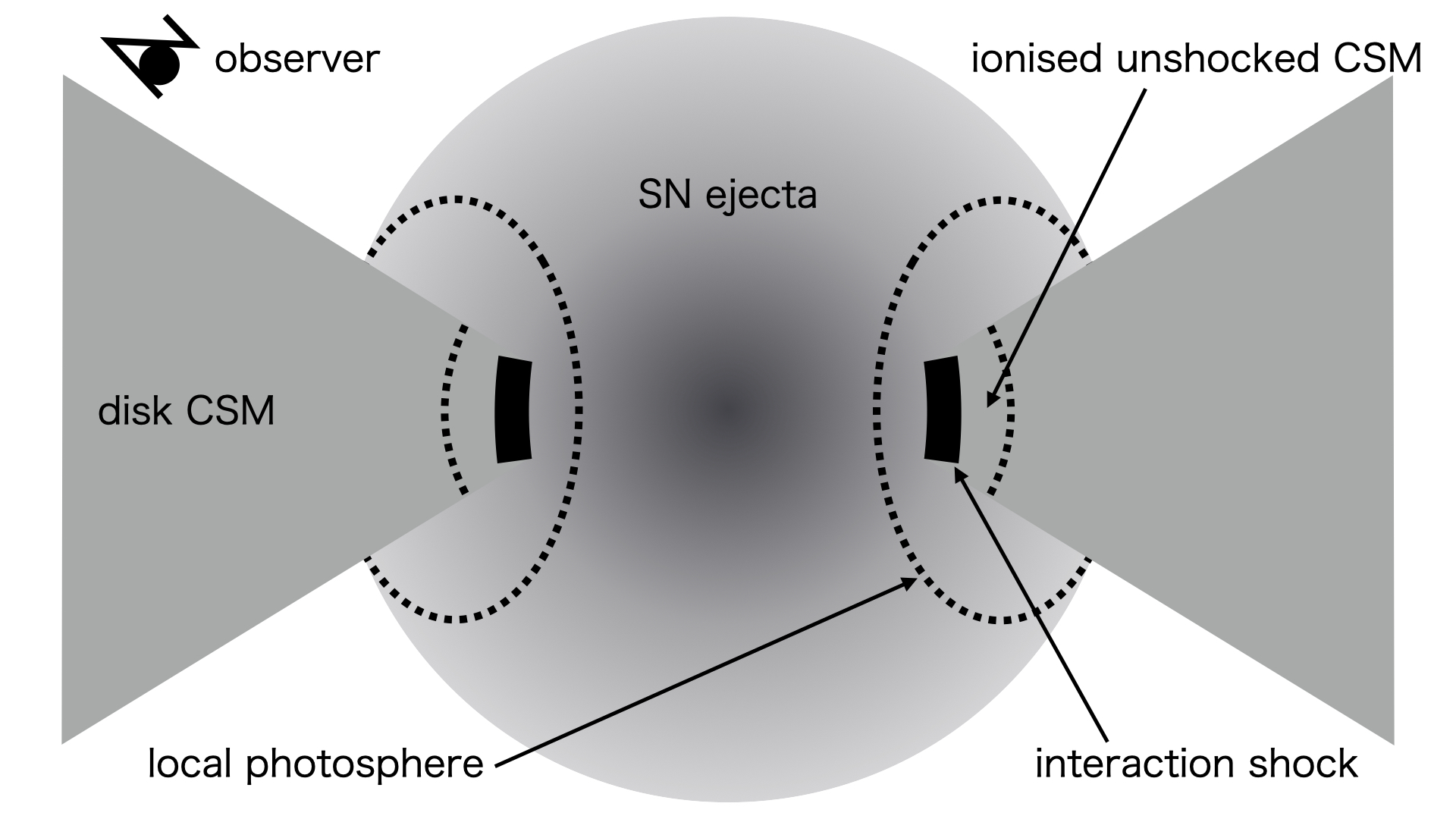}
      \caption{
      Schematic picture of the aspherical CSM interaction in SN~2021irp.
              }
         \label{fig:schematic_picture}
   \end{figure*}

Combining the above considerations with the observed persistent high continuum polarisation in SN~2021irp (see Section~\ref{sec:pol_properties}), we conclude that the radiation should mainly come from an aspherical (i.e., toroidal) photosphere and that this photosphere is located around the ring-shaped CSM interaction region, as shown in Figure~\ref{fig:schematic_picture}. 

\section{Discussion} \label{sec:discussion}

\subsection{Origin of the broad Balmer lines}

SN~2021irp notably shows strong broad (FWHM~$\sim8000$ km~s$^{-1}$ Balmer lines at least from 200~d after the explosion, rather than the narrow or intermediate width lines which are observed in Type~IIn SNe and regarded as a sign of strong interaction with H-rich CSM. This cannot be explained with the classical picture of Type~IIn SNe with a spherical CSM interaction \citep[e.g., Figure~1 in][]{Smith2017}, and thus provides an opportunity to improve our understanding of the radiation processes in an aspherical CSM interaction. In fact, the observed BB radius of SN~2021irp supports such an aspherical CSM interaction (see Paper~I).

\citet[][]{Smith2015} have proposed a qualitative picture for the photometric and spectroscopic behaviour of interacting Type~II SNe with a disc CSM in order to explain the observations of PTF11iqb \citep[see also][and references therein]{Smith2017}. At early phases, the CSM interaction region, which lies in the equatorial plane, is hidden by the opaque SN ejecta in the polar directions. Therefore, in this phase, the system would look like a luminous Type II SN, where its spectrum will be similar to those of normal Type II SNe with broad lines but without narrow lines. Since the photosphere in the SN ejecta in the polar directions recedes inward with time, we eventually see the interaction region. At this point, we would observe it as an interacting SN with a spectrum with intermediate width lines arising from the post-shock CSM. \citet[][]{Nagao2020} quantitatively investigated this picture for Type II SNe interacting with a CSM disc by calculating the shock evolution and radiative transfer, and demonstrated that such a phase of hidden CSM interaction can last only for the first several tens of days. Therefore, the location of the CSM interaction should already be exposed outside of the photosphere in the SN ejecta in the polar directions at the epochs of our spectroscopic observations (Phases $\gtrsim 200$~d after the explosion), and we can naively expect to see narrow Balmer lines. However, SN~2021irp does not fit this picture, as it shows strong broad Balmer lines, and does not show strong narrow lines in its spectra (see Paper~I).

We propose a modified picture for the photometric and spectroscopic behaviour of interacting Type~II SNe with disk CSM, as shown in Figure~\ref{fig:schematic_picture}. After the interaction runs over the photosphere in the SN ejecta in the polar directions, the high-energy photons from the interaction shocks heat and re-ionise the already recombined hydrogen gas, and create an optically thick region locally around the interaction. This reprocesses the original radiation created in the interaction region into radiation from the local ``photosphere", located in the SN ejecta around the interaction region (continuum plus broad lines). 

Since the local photosphere at this phase is located in a relatively outer region in the SN ejecta and distorted into the CSM disk direction, the line-forming regions are placed in higher velocity parts of the SN ejecta and thus higher-velocity-parts-enhanced and emission-dominated Balmer lines are expected even at late phases. Notably, in this case, the photosphere is not spherical but it should have a torus-like shape. This might explain the peculiar shapes of Balmer lines in SN~2021irp, which deviate from a Gaussian (see Paper~I).

   \begin{figure}
   \centering
   \includegraphics[width=\columnwidth]{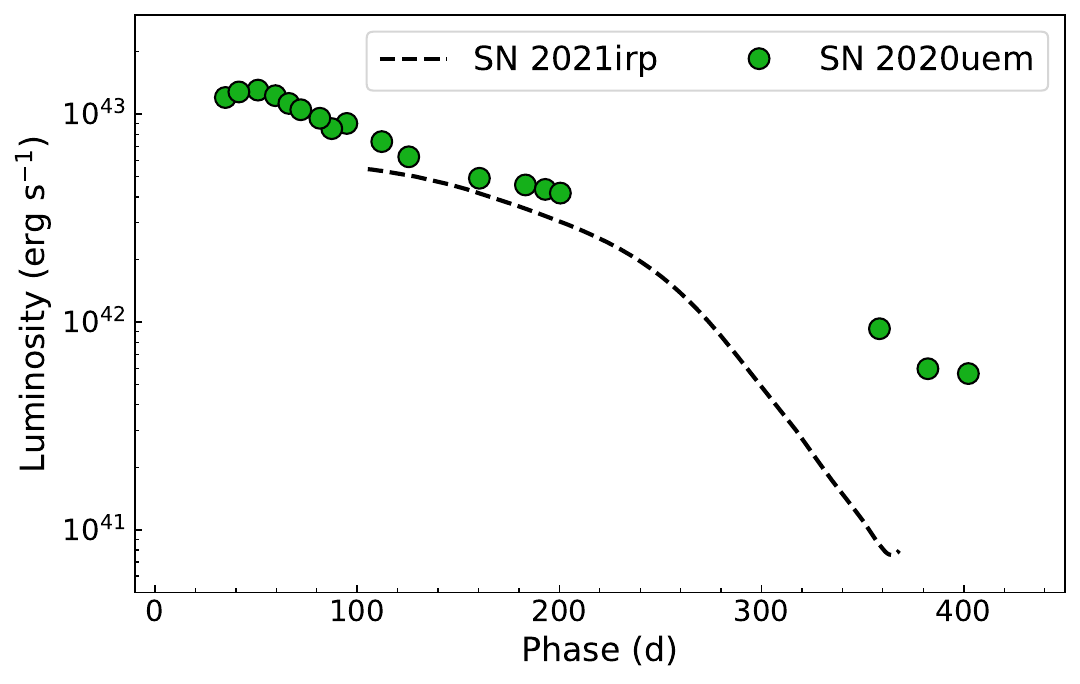}
      \caption{Pseudo-bolometric light curves of SN~2021irp and SN~2020uem. The luminosity is measured in the $B/g$-$I/i$ bands for both SNe.
              }
         \label{fig:LC_20uem}
   \end{figure}

   \begin{figure*}
   \centering
   \includegraphics[width=\textwidth]{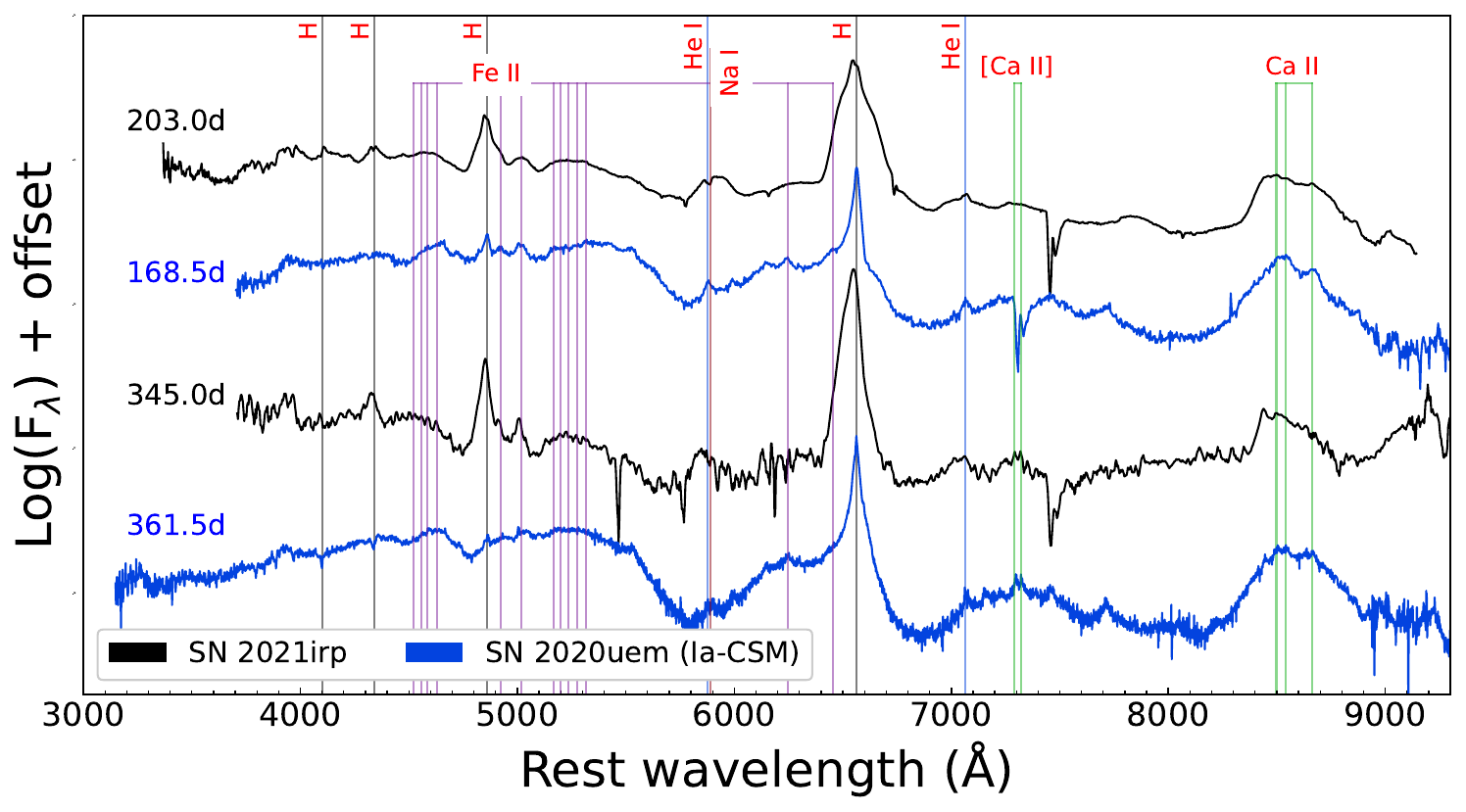}
      \caption{The spectra of SN~2021irp and SN~2020uem, compared at similar epochs. The labels give the phases the spectra were observed at. All spectra are corrected for extinction. 
              }
         \label{fig:spec_20uem}
   \end{figure*}

Comparing SN~2021irp with the Type Ia-CSM SN~2020uem can help support this interpretation. SN~2020uem is a Type Ia SN that has been interpreted as interacting with a similar CSM disk as we are suggesting for SN~2021irp, as implied by its photometric and polarimetric properties\citep[see][]{Uno2023a,Uno2023b}. As shown in Fig. \ref{fig:LC_20uem}, their bolometric light curves are relatively similar during the phases of SN~2021irp that are not strongly affected by increased extinction from newly formed dust. SN~2020uem also shows a high degree of polarisation (around 1\%), which is similar to SN~2021irp. Fig.~\ref{fig:spec_20uem} shows selected spectra of SN~2021irp and SN~2020uem at similar phases. There are a number of similar features in the spectra, particularly the \ion{Ca}{II} NIR triplet and \ion{Fe}{II} pseudo continua. However, SN~2020uem lacks the broad ($\sim$~8000~km~s$^{-1}$) emission line features of the Balmer series and \ion{He}{I}, instead showing narrow Balmer emission lines (1500 km~s$^{-1}$) with Lorentzian profiles at all times \citep{Uno2023a}.

Our interpretation of the origin of the broad Balmer lines in SN~2021irp naturally explains the different shapes of Balmer lines in the SNe~2021irp and 2020uem, both of which have similar configurations for the main energy source, i.e., interaction with several solar masses of CSM configured in a disk. In the case of SN~2021irp, the SN ejecta is expected to be hydrogen-rich and massive ($\sim 10$ M$_{\odot}$), while SN~2020uem has Type~Ia SN ejecta, which is hydrogen-poor and less massive. 
In the case of SN~2020uem, the SN ejecta around the shock regions cannot produce an optically thick photosphere hiding the interaction regions and the ionised un-shocked CSM due to the lower opacity in a Type Ia SN ejecta. Thus, we see the strong, narrow Balmer lines from the ionised unshocked CSM, unlike the case of SN~2021irp.

\subsection{Relations with other types of SNe}

Some interacting SNe, known as Type~IIn SNe, show strong narrow Balmer lines. These SNe may be explained with spherical CSM interaction as in the classical picture \citep[e.g.,][]{Smith2017}. However, if we observed the system of SN~2021irp from the equatorial plane of the CSM disk, we would also see strong narrow Balmer lines as in Type~IIn SNe. For example, SNe~2010jl, 2021acya, or 2021adxl, which all show strong multiple-width components of Balmer lines, might be such cases \citep[][]{Fransson2014,Brennan2024,Salmaso2024}. From the light curve modelling of SN~2021irp, the half-opening angle of the CSM disk is between $\sim30~$-$~90$\textdegree, while the high polarisation rules out very large opening angles ($\gtrsim 50$\textdegree). If we observe the system of SN~2021irp with random viewing angles with the assumption of a half-opening angle of 45 degrees, we would observe a 21irp-like SN (dominated by broad Balmer lines) for 30\% of all the viewing angles and a Type~IIn SNe for the remaining 70\%. Since we do not know the rate of 21irp-like SNe, it is not possible to estimate what fraction of Type~IIn SNe could originate from similar objects as SN~2021irp observed with edge-on views of the CSM disk.
Polarimetry is the key to unveiling SNe 
interacting with aspherical CSM that are viewed at a viewing angle that causes them to otherwise appear similar to Type~IIn that are dominated by spherical CSM interaction. Both types of transients would show narrow Balmer lines and thus should be classified as Type~IIn SNe. However, for the former case, we expect to have an aspherical photosphere and thus high polarisation, while we would get low polarisation for the latter case. Understanding the relative rates of each CSM configuration would reveal much about the origins (the progenitors and mass-loss mechanism) of Type~IIn SNe, and is therefore important to investigate.


We speculate that broad emission-dominated lines during the photospheric phases in other types of SNe (e.g., some Type IIL SNe, some Type~II superluminous SNe, and some Type II SNe as SN~2021irp) might share similar radiation processes. Indeed, the Type IIn SN 2017hcc shows very similar photometric, spectroscopic and polarimetric properties (see Paper~I). One possibility is that it is a similar system as SN~2021irp but with a different viewing angle. There might be more 21irp-like objects, observed with different viewing angles, within the population of Type IIn SNe and/or Type~II superluminous SNe.
Again, the key observation to identify 21irp-like SNe from the zoo of interacting SNe would be polarimetry, which directly traces the shape of the photosphere. In addition, the line profiles, width, and in particular their deviations from a Gaussian shape are useful for recognizing these hidden-interaction-powered SNe.

\subsection{The progenitor of SN~2021irp}

In this subsection, we discuss possible progenitors of SN~2021irp based on the estimated properties of its CSM. The light curve modelling of SN~2021irp demonstrates that this SN can be explained by a normal Type~II SN (i.e. an explosion of a $\sim 10$ M$_{\odot}$ red supergiant with an explosion energy of $\sim 10^{51}$ erg) interacting with disk CSM of $\gtrsim 2$~M$_{\odot}$, created within a few decades before the explosion (assuming a wind velocity of $\sim 50$ km s$^{-1}$). The polarimetric and spectroscopic properties of SN~2021irp also support this interpretation.
An interesting point here is that this peculiar SN with a very long duration does not require any peculiar progenitor star. It merely needs a normal massive star similar to those identified as progenitors of Type~IIP SNe \citep[e.g.][]{Smartt2009} but with an extremely large mass ejection just before the explosion. There is some observational evidence that supports this result: the progenitors of Type IIP and Type IIn SNe arise from similar environments \citep{Anderson2012}, and the locations of Type IIn SNe within their host galaxies do not follow the locations of the most intense star formation, as would be expected if they all arise from very massive stars \citep{Habergham2014,Ransome2022}. As for the origin of such massive CSM, any known mass-loss mechanisms in massive stars struggle to eject their outer parts in such a short period just before the explosion \citep[e.g.,][]{Smith2014b}, although some qualitative ideas for extensive mass ejections at the final evolutionary phases of massive stars have been proposed \citep[e.g.,][]{Humphreys1994,Langer1999,Yoon2010,Arnett2011,Chevalier2012,Quataert2012,Soker2013,Shiode2014,Smith2014a,Woosley2015,Quataert2016,Fuller2017}.

The observed properties of SN~2021irp, in particular the relatively symmetric shape of the H$\alpha$ emission line and the smooth light curve evolution, indicate that the CSM distribution is not clumpy or jet-like but a disk-like structure. This might suggest that the mass ejection process is related to binary interaction. Even with known types of mass ejections due to binary interaction, it is difficult to reproduce such an extreme mass ejection as we infer for SN~2021irp \citep[e.g.,][]{Ouchi2017}. One possibility is an extreme binary interaction, e.g., a common envelope evolution triggered by some unknown mechanism, closely followed by a SN \citep[see][for a possible example]{Pastorello2019}. Binary interaction is relatively common during the lifetime of progenitor systems of Type II SNe \citep[see e.g.][]{Zapartas2019}. However, since the timing of the mass ejection should be within a few decades of the explosion, the unknown mechanism that triggers such an extreme mass ejection should be related to the last burning phases (carbon and/or neon burning) in the progenitor star. For example, expansion of the stellar envelope related to the final burning phases could trigger binary interaction and thus enhance mass ejection during the last decades before the explosion (Murata \& Maeda, in prep). 

A suitable progenitor binary system for SN~2021irp should have a separation close enough to trigger this very late stage interaction, but far enough apart that there is no extensive history of mass transfer. A closer binary with extensive mass transfer would lead to a hydrogen-poor SN, while a more distant one with no mass transfer would not create the CSM we observe, and lead to a Type IIP SN. As SNe with such an extensive CSM interaction are rare \citep[Type IIn SNe make up 5-10\% of core-collapse SNe, see e.g.][]{Li2011,Smith2011,Cold2023}, and these progenitor systems would only be a subset of these SNe, such a fine tuned, and therefore rare, system is plausible.
It is important to increase the sample of this kind of SNe, and study the diversity in the parameters of the SN ejecta and CSM, in order to clarify their origin.

\section{Conclusions}

The polarimetric observations and light curve modelling presented in this work, supported by the observations and analysis presented in Paper I, show that the bright, long-lived Type~II SN~2021irp can be explained as a typical Type~II SN interacting with a CSM disk. The estimated mass and half-opening angle of the CSM disk are $\gtrsim 2$~M$_{\odot}$ (with corresponding mass-loss rate of $\sim0.035$ - $0.1$ M$_{\odot}$ yr$^{-1}$) and $\sim30-50$ \textdegree, respectively. We suggest that this CSM disk was created by some binary interaction process, and that SN~2021irp arises from a usual massive star (with Zero-Age-Main-Sequence mass of $\sim 8-18$ M$_{\odot}$). The separation and/or a mass ratio of the binary system should be within a specific range of values, that leads to a an extreme mass ejection in the final decades before the SN, but does not strip the star's H envelope at an earlier stage.

Studying SNe such as SN~2021irp is important for understanding not only their origin but also the origin of other interacting SNe, which may also be undergoing asymmetric interaction with a disk of CSM. We anticipate that there should be some Type~IIn SNe that have a similar CSM-disk interaction as SN~2021irp but with a different viewing angle (i.e., the edge-on view). Such Type~IIn SNe should have similar photometric evolution but different spectral features, in particular, strong narrow Balmer lines. We emphasize that polarimetry is important to distinguish narrow-line-dominated 21irp-like objects, powered by aspherical CSM interaction with edge-on views of the CSM disk, and Type IIn SNe that are powered by a spherical CSM interaction.

Based on the observational properties of SN~2021irp (and similar SNe), we also propose a new picture for the spectroscopic properties of Type II SNe interacting with a massive disk CSM. At early phases, they show a blue continuum and narrow Balmer lines. Once the interaction regions are hidden by the optically thick SN ejecta, the radiation from the interaction shock and the unshocked CSM is reprocessed, and they show a continuum and broad Balmer lines. When the SN ejecta become optically thin, they might start to show a minimal continuum and forbidden lines (so-called nebular spectrum), if the SN ejecta are still bright enough.

\begin{acknowledgements}

T.M.R is part of the Cosmic Dawn Center (DAWN), which is funded by the Danish National Research Foundation under grant DNRF140. T.M.R and S. Mattila acknowledge support from the Research Council of Finland project 350458. 
TN thanks Masaomi Tanaka and Akihiro Suzuki for fruitful discussion. TN and HK acknowledge support from the Research Council of Finland projects 324504, 328898, and 353019.
TK acknowledges support from the Research Council of Finland project 360274.
C.P.G. acknowledges financial support from the Secretary of Universities and Research (Government of Catalonia) and by the Horizon 2020 Research and Innovation Programme of the European Union under the Marie Sk\l{}odowska-Curie and the Beatriu de Pin\'os 2021 BP 00168 programme, the support from the Spanish Ministerio de Ciencia e Innovaci\'on (MCIN) and the Agencia Estatal de Investigaci\'on (AEI) 10.13039/501100011033 under the PID2023-151307NB-I00 SNNEXT project, from Centro Superior de Investigaciones Cient\'ificas (CSIC) under the PIE project 20215AT016 and the program Unidad de Excelencia Mar\'ia de Maeztu CEX2020-001058-M, and from the Departament de Recerca i Universitats de la Generalitat de Catalunya through the 2021-SGR-01270 grant. 
K.M. acknowledges support from the Japan Society for the Promotion of Science (JSPS) KAKENHI grant JP24KK0070 and 24H01810. The work is partly supported by the JSPS Open Partnership Bilateral Joint Research Projects between Japan and Finland (K.M. and H.K.; JPJSBP120229923).
M.F. is supported by a Royal Society - Science Foundation Ireland
University Research Fellowship.
N.E.R. acknowledges support from the PRIN-INAF 2022, `Shedding light on the nature of gap transients: from the observations to the models.
This research was partly based on observations made with the Nordic Optical Telescope (program ID P64-507 \& P64-023), owned in collaboration by the University of Turku and Aarhus University, and operated jointly by Aarhus University, the University of Turku and the University of Oslo, representing Denmark, Finland and Norway, the University of Iceland and Stockholm University at the Observatorio del Roque de los Muchachos, La Palma, Spain, of the Instituto de Astrofisica de Canarias.
The data presented here were obtained in part with ALFOSC, which is provided by the Instituto de Astrofisica de Andalucia (IAA) under a joint agreement with the University of Copenhagen and NOT.
This research was partly based on observations collected at the European Southern Observatory under ESO programmes 108.228K.001 and 108.228K.002.

\end{acknowledgements}

%

\bibliographystyle{aa} 
\bibliography{bibliography} 

\end{document}